\documentclass{ws-procs9x6}

\usepackage{verbatim} 

\begin{document}
\title{Local analysis of the sine-Gordon kink quantum fluctuations.}
\author{In\'es Cavero-Pel\'aez}
\address{Departamento de F\'isica Te\'orica. Universidad de Zaragoza, Spain.\\
E-mail:cavero@unizar.es, cavero@nhn.ou.edu}
\author{Juan Mateos Guilarte}
\address{Departamento de F\'isica Fundamental y IUFFyM. Universidad de Salamanca, Spain.\\
E-mail:guilarte@usal.es}
\begin{abstract}
We develop a local study of the Green's function and energy density for mesons fluctuating around the sine-Gordon kink background.
\end{abstract}
\keywords{Sine-Gordon kink, quantum fluctuations, Green's function, local energy density.}
\bodymatter
\section{Introduction.}
Topological defects are solutions of classical field equations that break translational and other symmetries. Finite energy solutions that conserve their shape at all times, even after collisions, are called solitons. In this paper, we shall study the fluctuations around a well known soliton, the sine-Gordon kink.\\
We consider a scalar field $\bar{\Phi}(t,\bar{z})$ in (1+1)
dimensions and Lagrangian dynamics given by
\begin{equation}
\bar{\mathcal{L}}=-\frac{1}{2}\partial_\mu\partial^\mu\bar{\Phi}^2-\bar{U}[\bar{\Phi}],
\end{equation}
with potential energy density,
\begin{equation}
\bar{U}[\bar{\Phi}]=\frac{m^4}{\lambda}(1-\cos\frac{\sqrt\lambda}{m}\bar{\Phi})\quad\rightarrow\quad
U [\Phi]=(1-\cos\Phi),
\end{equation}
where we have performed a change in scale to deal with
non-dimensional fields, space-time coordinates, and Lagrangian
density
\begin{equation}
\Phi=\frac{\sqrt\lambda}{m}\bar{\Phi},\qquad z=m\bar{z},\qquad \mathcal{L}=\frac{\lambda}{m^4}\bar{\mathcal{L}}.
\end{equation}
In this system the classical vacua associated to the minima of $U(\Phi)$ are infinitely degenerated, $\Phi_n=2\pi n,\, n\in\mathbb{Z}$. The choice of vacuum breaks the internal dihedral symmetry ${\mathbb D}_\infty={\mathbb Z}\times{\mathbb Z}_2=\{\Phi\to\Phi+2\pi n,\Phi\to -\Phi\}$ to the ${\mathbb Z}_2$ subgroup and there is also room for kink-solutions of the field equations connecting two of
such a vacua,
\begin{equation}
\Phi_K(t,z)=\pm4\arctan e^z+2\pi n.
\end{equation}
Studying quantum fluctuations around these kinks is a problem closely related to the analysis of vacuum fluctuations in the Casimir effect. In particular, the one-loop correction to the classical mass of the kink due to the kink quantum fluctuations can be seen as the kink Casimir effect, where the kink replaces the plates or other geometries constraining the dynamics of the fluctuations. Dashen, Hasslacher and Neveu \cite{DHN}{} (DHN) computed the sine-Gordon kink Casimir effect as long back as the mid seventies. The classical kink energy at the center of mass is $E[\Phi_K]=8m^3/\lambda$ (versus that of the vacua $E[\Phi_n]=0$) and DHN found the quantum correction to be $E_C[\Phi_K]=-m/\pi+\mathcal{O}(\lambda/m)$.\\
They controlled the ultraviolet divergences arising in this calculation by a combined regularization/renormalization procedure: 1) the kink and vacuum zero point energy were regularized using a cut-off in the number of normal fluctuation modes and subsequently the vacuum zero point energy was subtracted from the kink zero point energy. 2) The result was still divergent and it was necessary to implement a mass renormalization process in both sectors of the field theoretical system.\\
Other spectral zeta function regularization processes closer in spirit to the techniques used in the theoretical study of the ideal Casimir effect were developed \cite{BMM}{} and successfully applied to the calculation of the one-loop mass correction to the sine-Gordon and other relativistic kinks \cite{AMAJW}{}.\\
The above scenario offers a global knowledge of the total amount of energy taken from the kink by the quantum fluctuations. Our aim is to attack the problem by studying the Green's function of the small fluctuations in the kink background by importing the Green's function method as it is used in the Casimir effect \cite{M}. This point of view has not been explored as extensively as the zeta function approach. We address here a concrete example in a formulation within the general framework developed in Bordag\&Lindig\cite{bordag-lindig} for generic field backgrounds, a paper focusing in local properties of vacuum fluctuations. Our purpose is to benefit from the local information that the time-space Green's function can give through the computation of the energy momentum tensor and the subsequent knowledge of the one-loop energy density and radiation pressure.
\section{Kink fluctuation dynamics.}
By expanding the classical field around the kink background  $\Phi(t,z)=\Phi(z)_K+\phi(t,z)$, the Lagrangian up to quadratic terms in the
quantum fluctuations $\phi(t,z)$ becomes,
\begin{equation}
\mathcal{L}=-\frac{1}{2}\partial_\mu\partial^\mu\phi^2-\frac{1}{2}V(z)\phi^2-
\frac{1}{2}\phi^2+ {\cal O}(\phi^3), \label{qlag}
\end{equation}
where the \lq\lq space dependent" mass is,
\begin{equation}
V(z)=\left\{\begin{array}{cc}
0,&z<-a,\\
-\frac{2}{\rm cosh^2 z},&-a<z<a,\\
0,&z>a.\end{array}\right. \label{potential}
\end{equation}
Here $a$ is a point on the z-axis where the effect of the kink is almost negligible, that means $\cosh^2a\rightarrow\infty$, which can as well be used as infrared cut-off. We are allowed to split the potential in this way thanks to the localization of the classical kink energy density.\\
Therefore, we face a three region problem where we will impose matching and boundary conditions. In a final step, we shall make $a$ tend to infinity. The four dimensional Green's function $G(x,x')$ is the solution of the Euler-Lagrange
PDE of (\ref{qlag}) with unit source,
\begin{equation}
\big [\partial_\mu\partial^\mu-V(z)-1\big ]\,G(x,x')=\delta(x-x').
\end{equation}
Notice that even though we deal with the $(1+1)$-dimensional sine-Gordon model, it can be easily embedded in the analogous system in $(3+1)$-dimensional Minkowski space by transforming the kink in a domain wall invariant on the x and y axes. Thus, we can Fourier-transform the Green's function in the perpendicular coordinates to the direction of the kink and on time,
\begin{equation}
G(x,x')=\int\frac{d\omega}{2\pi}\int\frac{d^2k_\perp}{(2\pi)^2}\, e^{-i\omega(t-t')}\, e^{i{\bf k_\perp}({\bf x_\perp}-{\bf x'_\perp})}\,g(z,z'). \label{FT}
\end{equation}
The reduced Green's function $g(z,z')$ satisfies,
\begin{equation}
\big [-\partial_z^2+\lambda^2+V(z)\big]\,g(z,z')=\delta(z-z'),\label{reduced-g}
\end{equation}
and the constant $\lambda$ is\footnote{In this paper, whenever we need to use the value of $\lambda$, we assume that ${\bf k_\perp=0}$ but the generalization is interesting since it is compulsory in the membrane or domain wall interpretation.}, $\lambda^2=k_\perp^2-\omega^2+1$.\\
We next solve the differential equation (\ref{reduced-g}) in the three regions determined by the potential in (\ref{potential}) for appropriate boundary conditions.
\section{Decaying boundary conditions.}
We assume continuity of the Green's function and its derivatives in $-a$ and $a$ and exponentially decaying solutions at infinity.\\
For $z<-a$ and $z>a$ we find,
\begin{equation}
g_1(z,z')=-\frac{1}{2\lambda}\,e^{\lambda(z_<-z_>)}+a_1\,e^{\lambda(z+z')},
\end{equation}
where $z_<$ and $z_>$ are the lesser or the greater of $z$ and $z'$. The coefficient is given by
\begin{equation}
a_1=\frac{e^{2\lambda a}}{2\lambda\Delta}\Big\{e^{-2\lambda a}\big[2\lambda(\lambda+\tanh a)\cosh^2 a-1\big]-e^{2\lambda a}\big[2\lambda(\lambda-\tanh a)\cosh^2 a-1\big]\Big\},
\end{equation}
with
\begin{equation}
\Delta=e^{-2\lambda a}-e^{2\lambda a}\,\delta^2,\qquad \delta=\big[2\lambda(\lambda-\tanh a)\cosh^2 a-1\big].
\end{equation}
In the region of the kink, where $-a<z<a$, we find the following Green's function,
\begin{eqnarray}
g_2(z,z')&=&-\frac{1}{2\lambda}\frac{1}{(-\lambda^2+1)}\Bigg\{ f_1^-(z_>)f_1^+(z_<)+\frac{\delta}{\Delta}\Big\{\delta \,e^{2\lambda a}\Big[f_1^+(z')\,f_1^-(z)+\nonumber\\
&&f_1^-(z')\,f_1^+(z)\Big]+\Big[f_1^+(z')\,f_1^+(z)+f_1^-(z')\,f_1^-(z)\Big]\Big\}\Bigg\},\label{g22}
\end{eqnarray}
where
\begin{subequations}
\begin{eqnarray}
f_1^-(z)=e^{\lambda z}(-\lambda+\tanh z)\\
f_1^+(z)=e^{-\lambda z}(\lambda+\tanh z).
\end{eqnarray}
\end{subequations}
The poles of $g_2(z,z')$ lay on $\omega=0$ and $\omega=\pm 1$ as they should since, in the limit $a\to\infty$, the potential in (\ref{potential}) is a transparent P$\ddot{\rm o}$sch-Teller well. When it is shifted by $1$, there is a bound state of zero energy $\omega=0$, the so-called translational mode, and a ``half-bound" state lying in the threshold of the continuous spectrum $\omega^2=1$; see Barton\cite{Bar}{} and Graham \& Jaffe\cite{Jaf}. Notice that $g_2(z,z')$ captures all these states whereas $g_1(z,z')$ does not include the kink translational mode. The decaying boundary conditions exclude the states in the continuous spectrum.\\
We can now construct the energy-momentum tensor and get the zero-zero component corresponding to the local density energy. In the area where the kink lives it has the form,
\begin{equation}
\langle T_{00}(z)\rangle=\frac{1}{2 i}\Big[\partial_z\partial_{z'}+\lambda^2+2\omega^2-\frac{2}{\cosh^z}\Big]g_2(z,z')\Big|_{z=z'}.
\end{equation}
Using equation (\ref{g22}) and after subtracting the bulk term we find,
{\small
\begin{eqnarray}
\langle T_{00}(z)\rangle&=&\frac{-1}{2i\lambda(-\lambda^2+1)}\frac{\delta}{\Delta}\Bigg\{\delta e^{2\lambda a}\Bigg[2\omega^2(\tanh^2z-\lambda^2)+\frac{1}{\cosh^2z}-3\frac{\tanh^2z}{\cosh^2z}\Bigg]\nonumber\\
&+&\cosh(2\lambda z)\Bigg[2(\omega^2+\lambda^2)(\lambda^2+\tanh^2z)+\frac{1}{\cosh^2z}(1-4\lambda^2-3\tanh^2z)\Bigg]\nonumber\\
&+&\sinh(2\lambda z)\Bigg[-4\lambda(\lambda^2+\omega^2)\tanh z+6\lambda\frac{\tanh z}{\cosh^2 z}\Bigg]\Bigg\}.
\end{eqnarray}}
\section{Periodic boundary conditions.}
Here, we only consider the region where the potential is different from zero and impose periodicity on the Green's function and its first derivative at the borders $-a$ and $a$. The resulting Green's function is then,
\begin{eqnarray}
g(z,z')&=&\frac{1}{2\lambda(-\lambda^2+1)}\Bigg\{f_1^-(z_<)f_1^+(z_>)-\frac{1}{\Delta^C}\Big\{\frac{\tanh a}{\Delta^0}\nonumber\\
&\times&\Big\{\Big[f_1^+(z')\,f_1^+(z)+f_1^-(z')\,f_1^-(z)\Big]+\Big[f_1^+(z')\,f_1^-(z)\nonumber\\
&+&f_1^-(z')\,f_1^+(z)\Big]\Big\}\Big\}+e^{-\lambda a}\Big[-\lambda(\lambda+\tanh a)\cosh^2a+1\Big]\nonumber\\
&\times&\Big[f_1^+(z')\,f_1^-(z)+f_1^-(z')\,f_1^+(z)\Big]\Bigg\},
\end{eqnarray}
where the deltas in the denominators are,
{\small
\begin{subequations}
\begin{eqnarray}
\Delta^0&=&e^{-\lambda a}(\lambda+\tanh a)-e^{\lambda a}(\lambda-\tanh a).\\
\Delta^C&=&e^{\lambda a}\Big[-\lambda(\lambda-\tanh a)\cosh^2a+1\Big]-e^{-\lambda a}\Big[-\lambda(\lambda+\tanh a)\cosh^2a+1\Big],\nonumber\\
\end{eqnarray}
\end{subequations}}
The zeros of $\Delta^0$ are at $\lambda=\pm 1$ whereas the zeroes of $\Delta^C$
occur for $\lambda=i\frac{\pi n}{a}-\frac{1}{a}\rm{arcth}(1/\lambda)$. Thus, we find the whole spectrum of meson fluctuations around the kink at the poles of the Green's function; $\omega=0$ and
$\omega^2=\frac{\pi^2}{a^2}n^2+1$ plus a term which is suppressed for large $a$. A very subtle point is the fact that the periodic boundary conditions allow a second zero mode, besides the normalizable one, that will disappear in the limit of $a\to\infty$.\\
The local density energy for a fixed $\omega$ is
\begin{eqnarray}
\langle T_{00}(z)\rangle&=&\frac{1}{4i\lambda(-\lambda^2+1)\Delta^C}\Bigg\{-\frac{\tanh a}{\Delta^0}\Big\{e^{-2\lambda z}\Big[2(\lambda^2+\omega^2)(\lambda+\tanh z)^2\nonumber\\
&+&\frac{1}{\cosh^2z}(-4\lambda^2+1-6\lambda\tanh z-3\tanh^2z)\Big]+e^{2\lambda z}\Big[2(\lambda^2+\omega^2)\nonumber\\
&\times&(-\lambda+\tanh z)^2+\frac{1}{\cosh^2z}(-4\lambda^2+1+6\lambda\tanh z-3\tanh^2z)\Big]\nonumber\\
&+&2\Big[2\omega^2(-\lambda^2+\tanh^2z)+\frac{2}{\cosh^2z}(1-3\tanh^2z)\Big]\Big\}\nonumber\\
&+&e^{-\lambda a}\Big[-\lambda(\lambda+\tanh a)\cosh^2a+1\Big]\Bigg[4\omega^2(-\lambda^2+\tanh^2z)\nonumber\\
&+&\frac{2}{\cosh^2z}(1-3\tanh^2z)\Bigg]\Bigg\}.
\end{eqnarray}
\section{Comparison and prospects.}
In order to calculate the energy, we need to integrate the above expressions over the whole region as well as over frequencies. After integration we are going to make $a\rightarrow\infty$, therefore we need to consider only the region where the kink lives, that means the interval $[-a,a]$. The total energy would be,
\begin{equation}
T_{00}=\frac{1}{2\pi}\int_{-\infty}^\infty\int_{-a}^ad\zeta\,dz\langle T_{00}(z,\zeta)\rangle,
\end{equation}
where we have rotated to the imaginary axes, $\omega\rightarrow i\,\zeta$.
\begin{figure}
\begin{center}
\begin{tabular}{ll}
\psfig{file=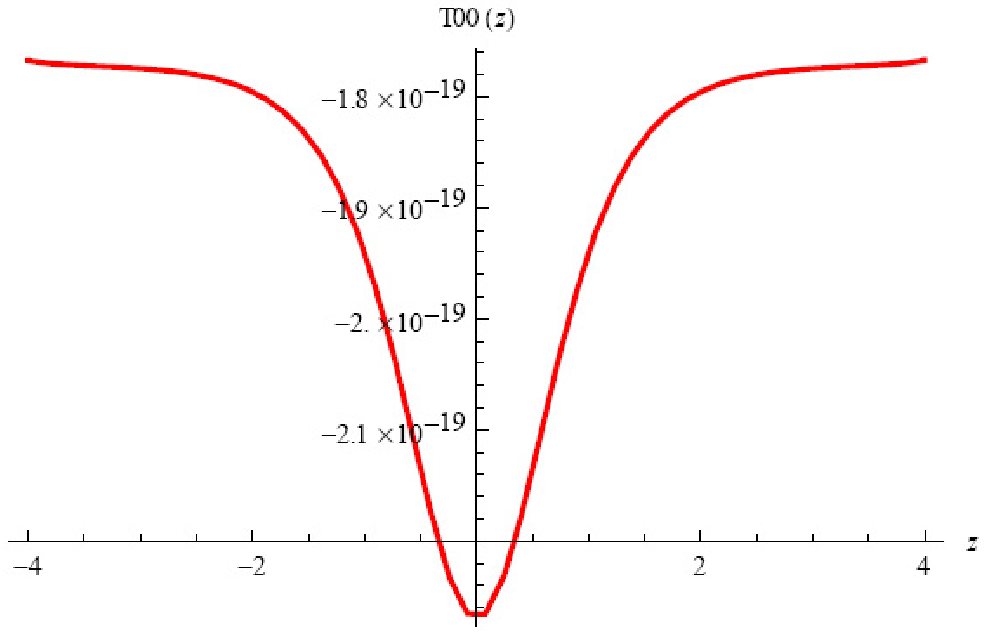,width=2.1in}
&\psfig{file=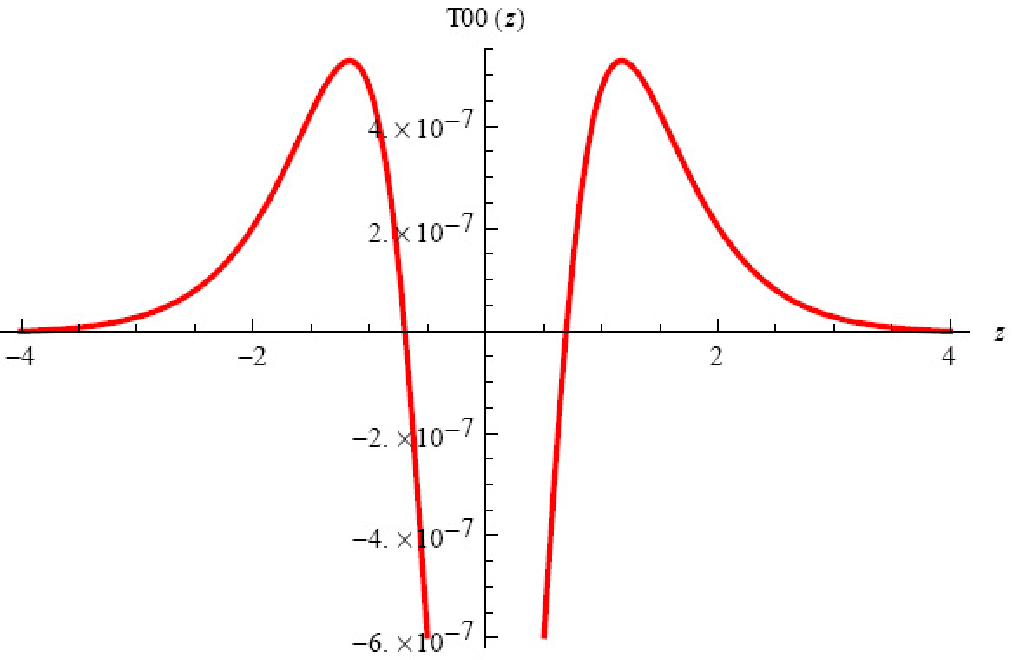,width=2.1in}
\end{tabular}
\end{center}
\caption{Local density energy for periodic boundary conditions. On the left $\omega=2$ and on the right $\omega=0.2$. The minimum in the right plot is $T_{00}=-1.87\times 10^{-6}$.}
\label{Cont_Priodc}
\end{figure}
Since we are interested in studying local effects, we focus on the integrand of the above expression and plot it for a large value of $a$ keeping $\omega$ constant (or equivalently $\zeta$) in each graph. In figure (\ref{Cont_Priodc}) we plot the local density energy around the kink for periodic boundary conditions for frequencies of $2$ on the left and $0.2$ on the right. The case of decaying boundary conditions at that frequency is qualitatively the same as for periodic, it differs on the value of the energy which is roughly the same as the plotted one except for the powers of ten which do not show for decaying boundary conditions. We observe by looking at the figure (\ref{Cont_Priodc}) that the behavior at low frequencies is quite different than that at large frequencies.
Clearly the effect of the kink over the passing mesons is stronger when they come with low frequency (graph on the right). The scale on the plots indicate that high frequency mesons passing by the kink background (graph on the left) are almost ignoring its presence and the bend on energy in that region is very small (notice that the plots do not have the same scale).
\begin{figure}
\begin{center}
\psfig{file=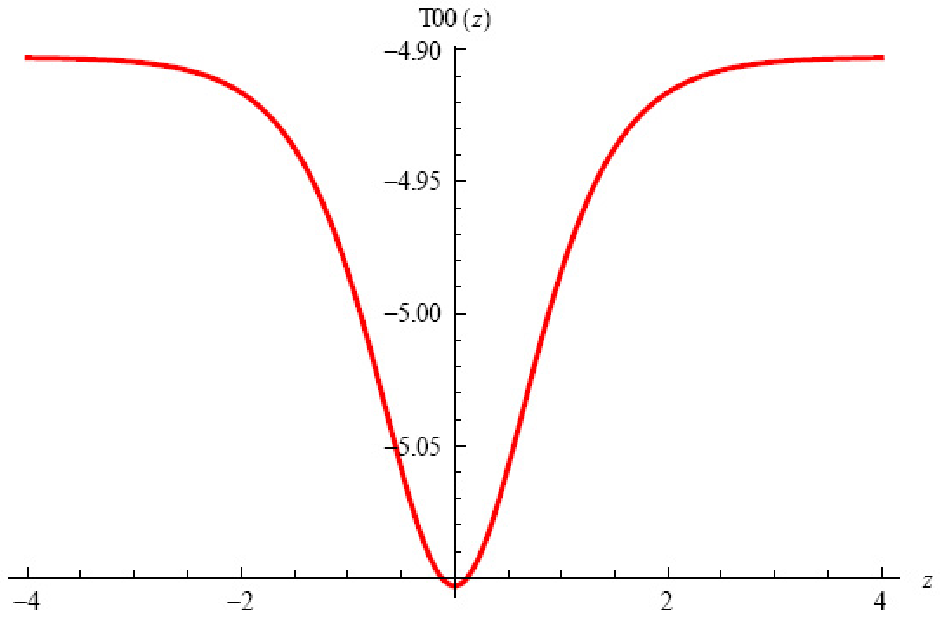,width=2.1in}
\begin{tabular}{ll}
\psfig{file=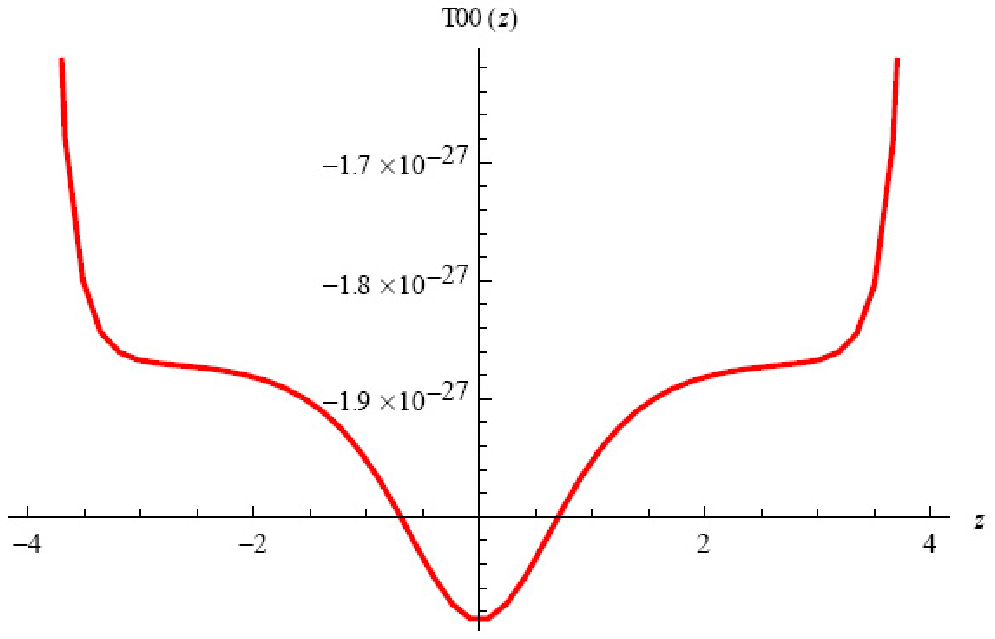,width=2.1in}
&\psfig{file=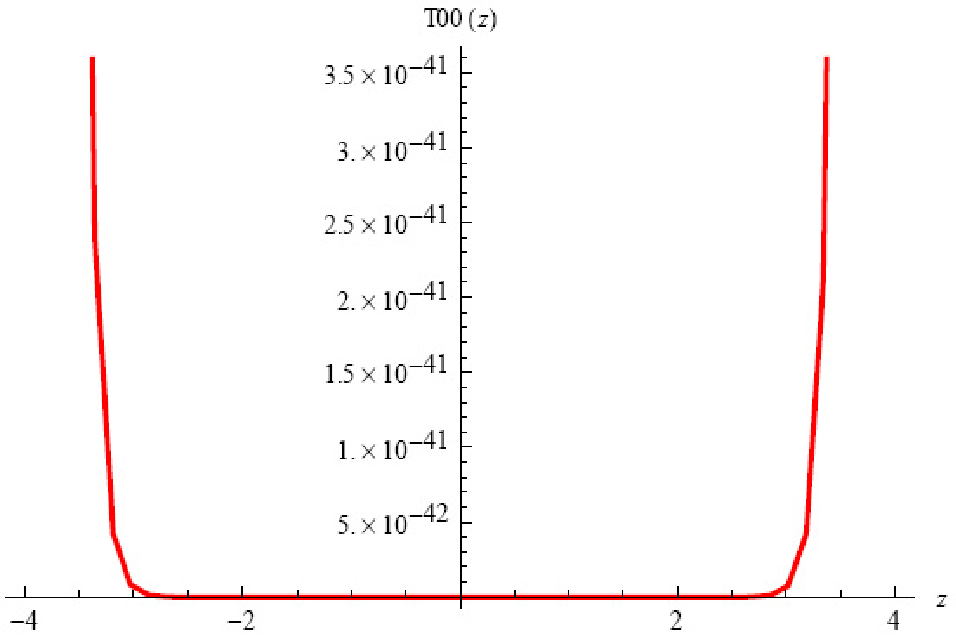,width=2.1in}
\end{tabular}
\end{center}
\caption{Density energy for high frequencies. Decaying boundary conditions on the top ($\omega=5$) and periodic boundary conditions on the bottom ($\omega=3$ left and $\omega=5$ right).}
\label{High_ws}
\end{figure}
Graphs shown in figure (\ref{High_ws}) describe the behavior for large frequencies. We observe that for decaying boundary  conditions (on the top), the value of the density energy tends to a constant value, while for periodic boundary conditions (on the bottom) the plot gets distorted as the frequency gets larger, and it diverges.\\
In a coming publication we hope to expand this analysis by regularizing and renormalizing the ultraviolet divergences following the heat kernel regularization method \cite{bordag-lindig} .
{\small
\section*{Acknowledgment}
We thank the European Science Foundation(ESF) within the activity `New Trends and Applications of the Casimir Effect' exchange grant 2301, the Spanish Ministerio de Educaci\'on y Ciencia and the Junta de Castilla y Le\'on under grants FIS2006-09417 and GR224 for partial support of this research. We especially thank Kimball A. Milton for an excellent organization and a successful workshop. IC acknowledges travel support by the ESF.}
\bibliographystyle{ws-proc9x6}
\bibliography{ws-pro-sample}

\begin{thebibliography}{99}
\addcontentsline{toc}{section}{References}

\bibitem{DHN} R.~Dashen, B.~Hasslacher, A.~Neveu, Phys. Rev. {\bf D10} (1974) 4130.
\bibitem{BMM} M.~Bordag, U.~Mohideen, V.~M.~Mostepanenko, Phys. Rept.{\bf 353}:1-2005,2001.
\bibitem{AMAJW} A.~A.~Izquierdo, W.~G.~Fuertes, M.~A.~G.~Leon, J.~M.~Guilarte, Nucl. Phys. {\bf B 635}(2002) 525-557.
\bibitem{M} K.~A.~Milton, J. of Phys {\bf A37}(2004)R209.
\bibitem{bordag-lindig} M.~Bordag and J.~Lindig, J. of Phys {\bf A29} (1996) 4481.
\bibitem{Bar} G.~Barton, J. of Phys {\bf A18}(1985)479.
\bibitem{Jaf} N.~Graham and R. Jaffe, Nucl. Phys, {\bf B544}(1999)432-447.

\end{thebibliography}

\end{document}